\documentclass[prl,aps,twoside,showpacs,superscriptaddress,twocolumn,floatfix,reprint]{revtex4}

\usepackage{amsmath}
\usepackage{latexsym}
\usepackage{epstopdf}
\usepackage{graphicx}
\usepackage{ulem}

\def\extra#1{{}}

\newcommand\ket[1]{\ensuremath{|#1\rangle}}
\newcommand\bra[1]{\ensuremath{\langle#1|}}

\def\extra#1{{}}
\begin{document}

\title{Experimental linear-optical implementation of a multifunctional optimal qubit cloner}

\author{Karel Lemr}
\affiliation{RCPTM, Joint Laboratory of Optics of Palack\'y
University and Institute of Physics of Academy of Sciences of
the Czech Republic, Faculty of Science, Palack\'y University
\\17. listopadu 12, 771~46 Olomouc, Czech Republic}
\author{Karol Bartkiewicz}
\affiliation{Faculty of Physics, Adam Mickiewicz University,
PL-61-614 Pozna\'n, Poland}
\author{Anton\'in \v{C}ernoch}
\affiliation{RCPTM, Joint Laboratory of Optics of Palack\'y
University and Institute of Physics of Academy of Sciences of
the Czech Republic, Faculty of Science, Palack\'y University
\\17. listopadu 12, 771~46 Olomouc, Czech Republic}
\author{Jan Soubusta}
\affiliation{Institute of Physics of Academy of Science of
the Czech Republic, Joint Laboratory of Optics of PU and IP
AS CR, 17. listopadu 12, 77207 Olomouc, Czech Republic}
\author{Adam Miranowicz}
\affiliation{Faculty of Physics, Adam Mickiewicz University,
PL-61-614 Pozna\'n, Poland}

\begin{abstract}

We present the first experimental implementation of a multifunctional device for the
optimal cloning of one to two qubits. Previous implementations have always been designed
to optimize the cloning procedure with respect to one single type of {\it a priori}
information about the cloned state. In contrast, our “all-in-one” implementation is
optimal for several prominent regimes such as universal cloning, phase-covariant 
cloning, and also the first ever realized mirror phase-covariant cloning, when the
square of the expected value of Pauli’s $Z$ operator is known in advance.In all these
regimes the experimental device yields clones with almost maximum achievable aver-
age fidelity (97.5\% of theoretical limit). Our device has a wide range of possible
applications in quantum information processing, especially in quantum communication.
For instance, one can use it for incoherent and coherent attacks against a variety of 
cryptographic protocols , including the Bennett-Brassard 1984 protocol of quantum
key distribution through the Pauli damping channels. It can be also applied as a 
state-dependent photon multiplier in practical quantum networks.
\end{abstract}


\maketitle

\noindent\textit{Introduction. }
One of the most fundamental laws of nature, the so-called no-cloning theorem, states
that an unknown quantum state cannot be perfectly copied. This fact has an imminent
impact on quantum information processing. For instance, it allows designing inherently
secure cryptographic protocols~\cite{cryptography1} or assures the impossibility of
superluminal communication~\cite{Bruss11}. Although perfect quantum copying is
impossible, one can still investigate how well such an operation can be approximated
within the limits of physical laws. Despite some very intense research in this domain,
many aspects of state-dependent quantum cloning have not yet been fully investigated.

Quantum cloning is one of the most intriguing topics in quantum physics. It is important
not only because of its fundamental nature but also because of its immediate
applications to quantum communications, including quantum cryptography. Similar
to other important quantum information processing protocols, quantum cloning has
undergone considerable development over the past two decades. The first design of an
optimal cloning machine was suggested by Bu\v{z}ek and Hillery~\cite{Buzek96}.
The cloner is called optimal when it gives the best results allowed by quantum
mechanics. Moreover universal cloning (UC) should operate equally well for all possible
qubit states~\cite{cloning1,cloning2, Ricci04,Irvine04,Khan04}. In contrast,
limiting cloning to a specific subset of qubit states, one can achieve a more precise
cloning operation. A prominent example of this situation is phase-covariant cloning
(PCC), where only qubit states with equal superposition of $|0\rangle$ and $|1\rangle$
are considered~\cite{cloning3,cloning4,cloning5, cloning6, Bruss00, DAriano03,F2}.

\begin{figure}[h!]
\includegraphics[scale=.3]{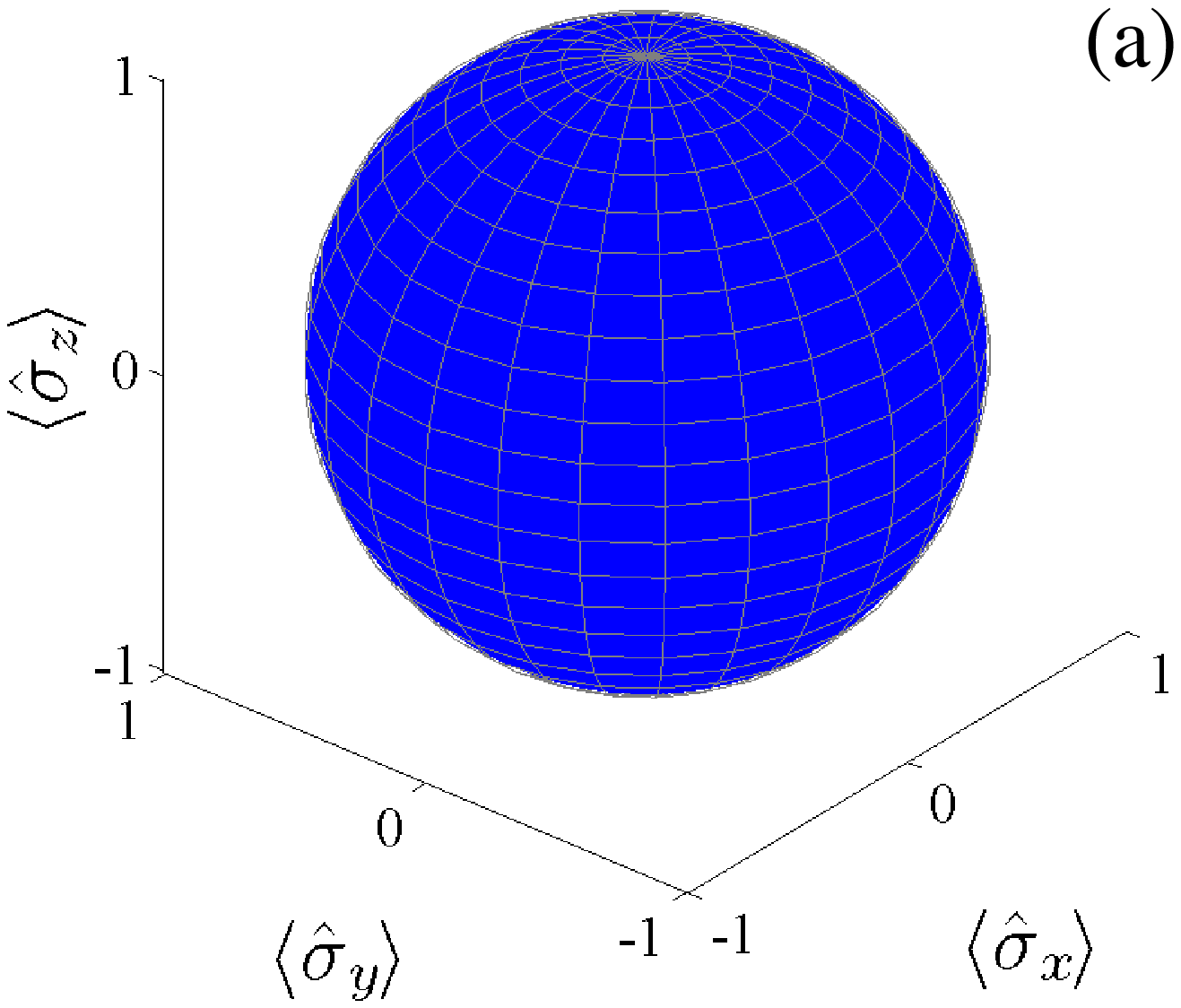}
\includegraphics[scale=.3]{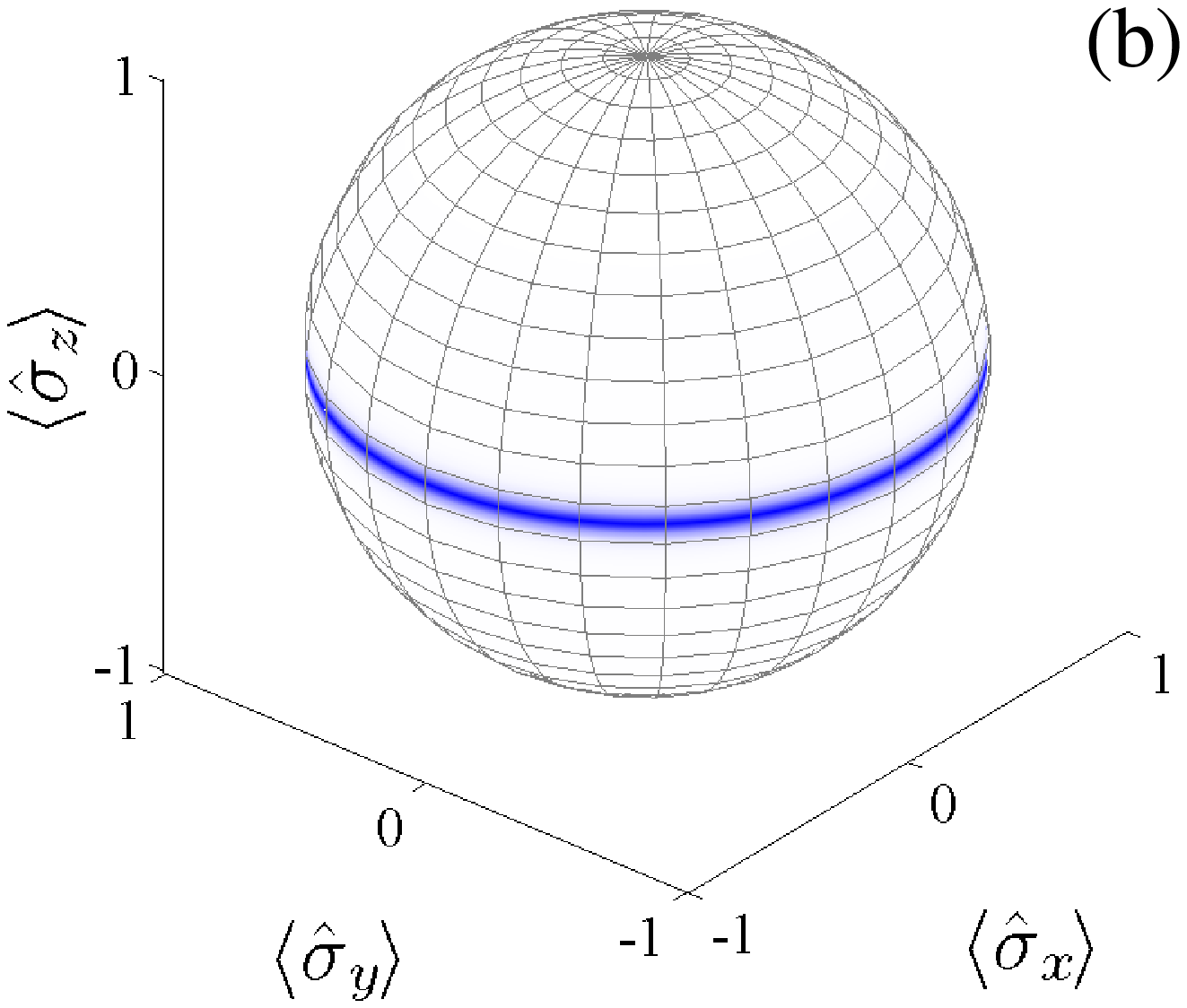}\\\vspace*{0.1cm}
\includegraphics[scale=.3]{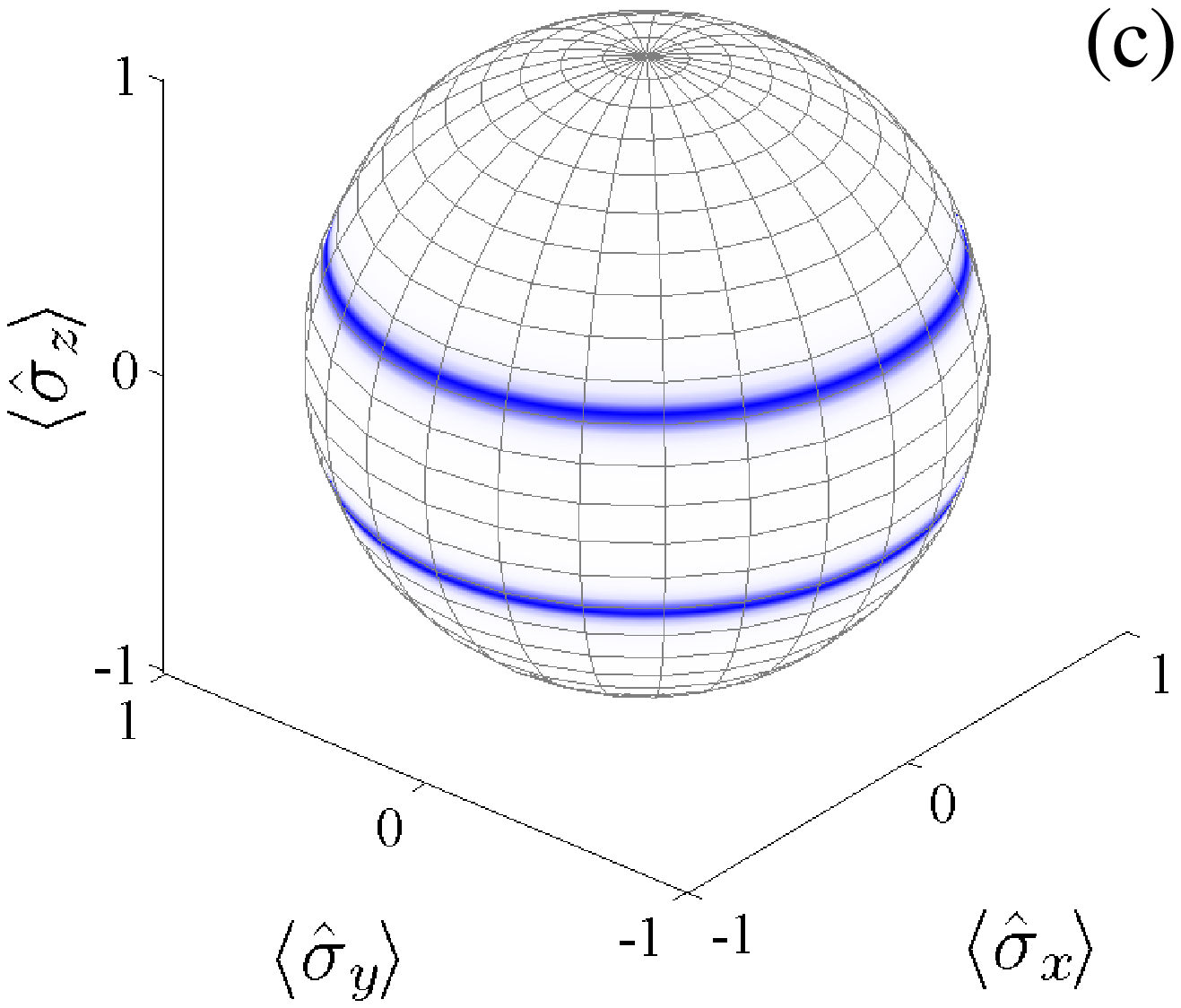}
\includegraphics[scale=.3]{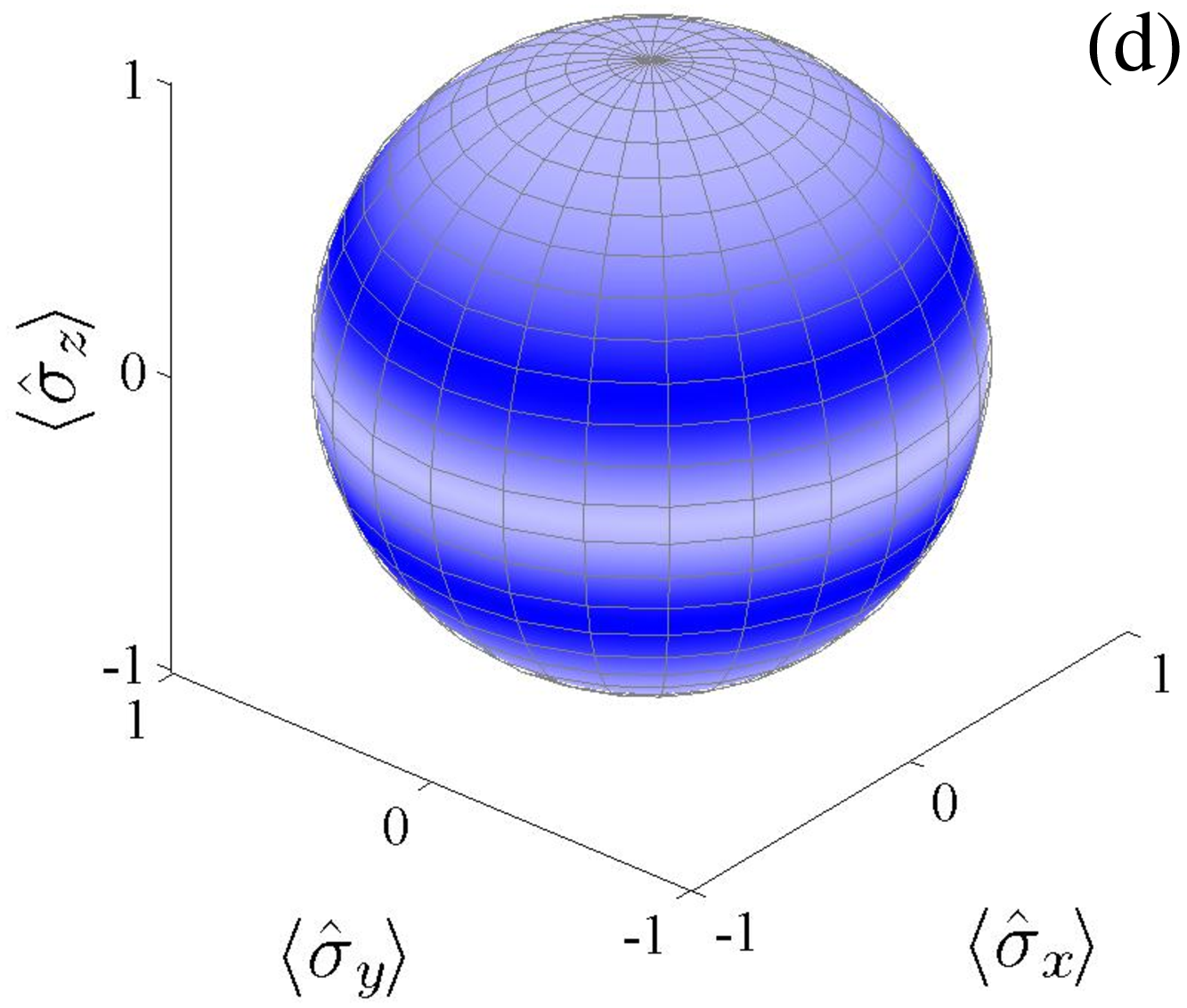}
    \caption{
(Color online) Overview of various distributions on Bloch's sphere
describing {\it a priori} knowledge of qubits and the corresponding
{\it optimal} cloning machines which are special cases of
our multifunctional cloner: (a) uniform distribution
can be cloned by the UC, (b) qubits on the
equator of the Bloch sphere can be cloned by the standard
PCC, (c) union of the set of qubits
for the generalized PCC and its equator-plane reflection can
be cloned by the MPCC, and (d) any set of qubits of unknown phase symmetric
about any equator plane can be cloned by the generalized
MPCC.
\label{fig:states}}
\end{figure}

In this Rapid Communication we address a question that is 
interesting from both conceptual and practical points of view:
how well a can quantum state be cloned if some {\it  a priori}
information about the state is known? 
Theoretical investigation of this issue led
to quantifying the information known about the cloned state
in terms of axially symmetric distributions on the Bloch
sphere~\cite{Bartkiewicz2}. This class of distributions
contains an important subclass of distributions which are
mirror symmetric with respect to the equatorial plane (see
Fig.~\ref{fig:states}). It is therefore convenient to
define {\it mirror phase-covariant cloning} (MPCC) as a
strategy for cloning states with this kind of {\it a priori}
information~\cite{Bartkiewicz}.

We hereby present an implementation of the MPCC, and we
also demonstrate that the same setup can be used for optimal
cloning in other prominent regimes such as universal cloning
and phase-covariant cloning. We show that the assumptions regarding the
symmetry of the set of qubits cloned in an optimal way by the
MPCC can be relaxed to include a wider class of qubit
distributions that do not need to be axially symmetric.
Finally, we demonstrate, for the example of the PCC for an
arbitrary polar angle on the Bloch sphere~\cite{F2}, that our
device can be also used as an optimal axially symmetric
cloner for which the mirror-symmetry condition is not
necessary.

\noindent\textit{Mirror phase-covariant cloning. } In our
experiment we cloned the polarization state of a single
photon given by
\begin{equation}
|{\psi}\rangle =
\cos(\theta/2)|H\rangle+\sin(\theta/2){\rm
e}^{i\varphi}|V\rangle, \label{eq:target:rho}
\end{equation}
where $|H\rangle$ and $|V\rangle$ are the horizontal and
vertical polarizations, respectively.  In accord with the
original definition~\cite{Bartkiewicz}, we assume
$\langle\hat\sigma_z\rangle^2 = \cos^2\theta_{\rm eff}$ is
the only {\it a priori} information known about the cloned
state, where $\hat\sigma_z$ denotes the third Pauli operator.
A geometrical interpretation of the set of states of fixed
$\cos^2\theta_{\rm eff}$ is shown in
Fig.~\ref{fig:states}(c). It has been recently
demonstrated~\cite{Bartkiewicz2} that the MPCC can also be
applied to a wider class of qubit distributions
$g(\theta,\varphi)$ shown in Fig.~\ref{fig:states}(d).
Consequently, the optimal cloner for a set of qubits given by
a distribution $g(\theta,\varphi)$ is an MPCC set for an
axial angle $\theta_{\rm eff}$ defined as
$\langle\cos^2\theta\rangle = \cos^2\theta_{\rm eff}$, where
the angle bracket stands for averaging over the distribution.
Moreover, we note that the mirror-symmetry condition can be
weakened and the MPCC transformation can be used as an
optimal cloning transformation for other sets of qubits which
are not axially symmetric and do not exhibit the
mirror-symmetry, but rather fulfill the following conditions:
\begin{eqnarray}
\nonumber
\int_{0}^{2\pi} [g(\theta,\varphi) +g(\pi-\theta,\varphi) ]{\rm e}^{i\varphi n}{\rm d}\varphi = 0,\\
\int_{0}^{2\pi} g(\theta,\varphi){\rm d}\varphi =
\int_{0}^{2\pi} g(\pi-\theta,\varphi){\rm d}\varphi,
\label{eq:weak_mpcc}
\end{eqnarray}
where $g$ is a distribution of qubits on the Bloch sphere and
$n=1,2$. Therefore, any MPCC optimal for some $\theta_{\rm
eff}$ is also optimal for a wider class of distributions
which do not need to be axially symmetric or
mirror symmetric but fulfill Eqs.~(\ref{eq:weak_mpcc}). The
above-mentioned arguments considerably broaden the usefulness
of the presented device.

\noindent\textit{Experimental setup. }
\begin{figure}
    \includegraphics[scale=0.33]{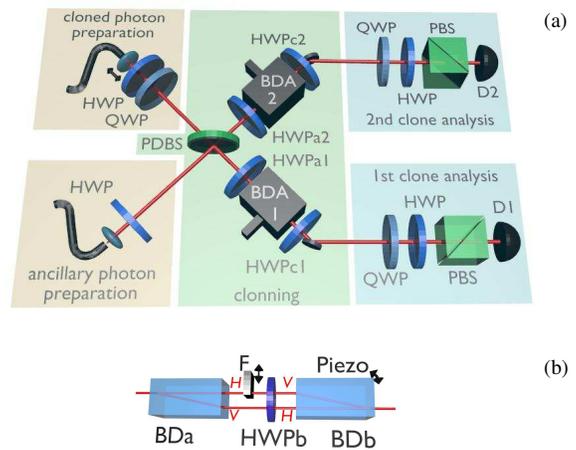}
\caption{\label{fig:setup} (Color online) (a) Scheme of the
experimental setup as described in the text. (b) Detailed
scheme of the beam-divider assembly. HWP, half-wave
plate; QWP, quarter wave-plate; PDBS, polarization
dependent beam splitter; BDA, beam divider assembly; F,
neutral density filter; PBS, polarizing beam splitter; D,
single-photon detector; BD, beam divider.}
\end{figure}
Our experimental setup is depicted in figure~\ref{fig:setup}.
First, the cloned and ancillary photon states are prepared by
means of half and quarter wave plates. Then the cloning
operation is performed by overlapping the two photons on
a special unbalanced polarization-dependent beam
splitter (PDBS). Subsequently, each of the two photons
undergoes polarization sensitive filtering (transmittance
$\tau$) using the beam divider assemblies (BDA1 and
BDA2) placed in each of the output modes of the beam
splitter. The PDBS employed in this scheme has different
transmittances for horizontal ($\mu$) and vertical ($\nu$)
polarizations. The transmittances should be given by
\begin{equation}
\mu = \frac{1}{2}\left(1+\frac{1}{\sqrt{3}}\right), \quad \nu
= \frac{1}{2}\left(1-\frac{1}{\sqrt{3}}\right),
\end{equation}
but due to manufacturing imperfections the observed
transmittances of our PDBS were $\mu = 0.76$ and $\nu =
0.18$. Please note that this imperfection can be corrected
without loss of fidelity through suitable filtering at
the expense of a lower success rate (see Ref.~\cite{supplement}).

The beam-divider assembly is depicted in more detail in
Fig.~\ref{fig:setup}(b). It is composed of two beam dividers
(BDa and BDb) used to separate and subsequently combine
horizontal and vertical polarizations. A neutral density
filter (F) with tunable transmittance $\tau$ is positioned
between the two beam dividers so that one of the paths
(polarizations) is attenuated while the other remains intact.
Also a half-wave plate (HWPb) is placed between the beam
dividers swapping the polarizations and thus allowing them to
be combined at the second beam divider (BDb). To control
attenuation of each polarization by the neutral density
filter, we envelope the beam-divider assembly by two
half-wave plates (HWPa and HWPc). The beam-divider assembly is
equivalent to a Mach-Zehnder interferometer, and by means
of a piezo-driven tilt of one of the beam dividers, we can
set an arbitrary phase shift between the two paths
(polarizations).

In the ideal case, having $\mu + \nu = 1,$ the setup operates
as follows. A separable two-photon state $|H_1H_2\rangle$
(indices denote the mode number) is generated in the process
of the type-I spontaneous parametric down conversion using a
LiIO$_3$ crystal pumped by cw Kr$^+$ laser at 413\,nm of
150-mW optical power. These photons are brought to the input
of the setup via single-mode fibers. The parameters to be set
for the PCC and UC regimes are just specific cases of the
MPCC setting as we discuss later. For this reason we now
concentrate on the MPCC setting. The polarization of the
first (cloned) photon is set in such a way that it belongs to
one of the parallels of latitude on the Bloch sphere with a
given polar angle $\theta$ [see Eq.~(\ref{eq:target:rho})].
The second (ancillary) photon remains either horizontally
polarized or is randomly swapped to vertical polarization.
After this preparation stage the two photons are coherently
overlapped at the PDBS. Depending on the polarization of the
ancillary photon, we perform subsequent transformation. If
the ancillary photon remains horizontally polarized we set
the half-wave plates (HWPa1 and HWPa2) in front of the beam
dividers to 45$^{\circ}$ so that the vertical polarization is
attenuated in both beam-divider assemblies. The level of
transmittance $\tau$ of the filters F is set according to the
relation
\begin{equation}
\tau =
\frac{\left({1-\Lambda^2}\right)\left(1-2\mu\right)^2}
{2\mu\nu\Lambda^2},\;\Lambda = \sqrt{\frac{1}{2} +
\frac{\cos^2\theta}{2\sqrt{P}}}, \label{eq:filter}
\end{equation}
where $P = 2-4\cos^2\theta+3\cos^4\theta$. Additionally, we also
set a phase shift $\pi$ between horizontal and vertical
polarization in both output modes. In the case of the
ancillary photon being vertically polarized we set the
half-wave plates HWPa1 and HWPa2 to 0$^{\circ}$ and this time
subject the horizontal polarization to the same filtering as
given by Eq.~(\ref{eq:filter}). Also we set the phase shift
between the polarizations to zero and rotate the half-wave
plates HWPc1 and HWPc2 to 45$^{\circ}$, thus canceling the
polarization swap exercised by the half-wave plates HWPb1 and
HWPb2 (inside the beam-divider assemblies).

Finally, the two-photon state polarization analysis is
carried out by measuring the rate of two-photon coincidences
for all combinations of single-photon projections to
horizontal, vertical, diagonal, anti diagonal linear, and
right and left circular polarizations \cite{Halenkova12}.
We can then estimate the two-photon density matrix using
a standard maximum likelihood method~\cite{Jezek03}.

In order to use the setup for the PCC, one just needs to set
all the parameters as if performing the MPCC set for the
latitude angle $\theta = \pi/2$. In case of the PCC there is
no need to randomly swap the horizontal and vertical
ancillae. In this case we know the hemisphere to which the
cloned states belong so we can simply use the closer ancilla
(horizontal for northern hemisphere and vertical for southern
hemisphere).

A similar analysis can be carried out to determine that the
setup actually performs the UC if set to the same parameters
as for the MPCC with the polar angle $\theta = {\rm
arccos}(\sqrt{3}/3)$.  In this regime a random swap between
horizontal and vertical ancillae is also required.

\begin{figure}
\includegraphics[scale=1]{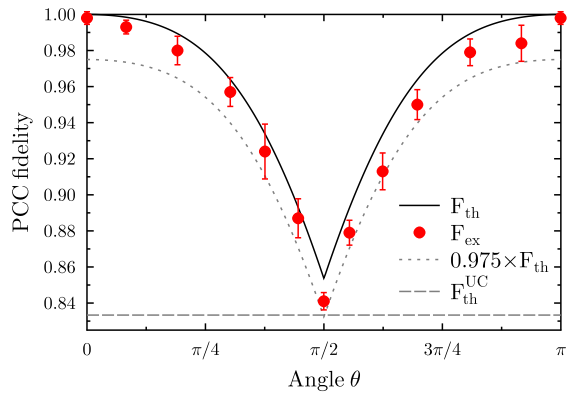}
\caption{\label{fig:pcc}(Color online) Experimental fidelity $F_{\rm ex}$
for the PCC regime depicted against theoretical prediction
$F_{\rm th}$ (solid curve). The short-dashed curve indicates
the fidelity drops to 97.5\% with respect to the corresponding
theoretical value. The theoretical fidelity $F_{\rm th}^{\rm
UC}$ for the UC is also depicted (long-dashed curve).
For the generalized PCC the hemisphere of the cloned qubit is 
known (the reverse is true for the MPCC); we choose the ancilla
deterministically~\cite{supplement} (horizontal for northern hemisphere and vertical for southern hemisphere), and we set transmittances of filters as for the
MPCC tuned for the angle $\theta=\pi/2$.
}
\end{figure}

\begin{figure}
\includegraphics[scale=1]{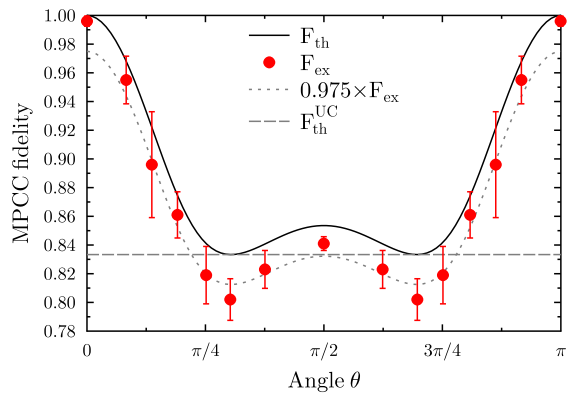}
\caption{\label{fig:mpcc}(Color online) Same as in Fig.~\ref{fig:pcc}, but
for the MPCC.}
\end{figure}

\begin{figure}
\includegraphics[scale=1]{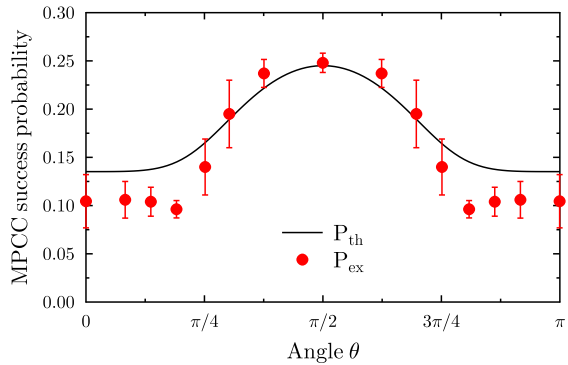}
\caption{\label{fig:mpcc_p}(Color online) The success probability of the
MPCC as a function of polar angle $\theta$:
$P_{\rm ex}$ denotes an experimentally determined value, and
$P_{\rm th}$ denotes our theoretical prediction. Note that
sometimes the experimental results surpass the theoretical
ones; this happens at the expense of lower fidelity of the
cloning process.}
\end{figure}

\noindent\textit{Experimental results. } In order to verify
the versatile nature of the cloner, we performed a series of
measurements in three regimes: PCC, MPCC, and UC. These
regimes differ just in the amount of {\it a~priori} knowledge
about the cloned state. For the PCC and MPCC we verified the
theoretical prediction of maximally achievable average
fidelity as a function of polar angle $\theta$. For all polar
angles (except the poles) we estimated the fidelities of both
clones for four different equally distributed input states.
The observed values are depicted in Fig.~\ref{fig:pcc} for
the PCC and similarly in Fig.~\ref{fig:mpcc} for the MPCC
regime. For
 UC (when we set $\tau = \tau[{\rm
arccos}(\sqrt{3}/3)]$) we cloned six input states: horizontally,
vertically, right and left circularly, diagonally and
anti diagonally polarized states. The average fidelity
obtained in the UC mode is 81.5$\pm$1.2\%. The vast majority of
the experimentally obtained fidelities in all regimes reached
or surpassed 97.5\% of their theoretical prediction leading
only to a very small experimental error.

Additional measurement of the success probability was performed
for the case of MPCC. The success probability as a function
of polar angle $\theta$ is depicted in Fig.~\ref{fig:mpcc_p}.
Note that success probability strongly depends on the
splitting ratio of the beam splitter. Its theoretical
prediction is given by
\begin{equation}
P_{\rm th} =  (1-2\mu)^2/2 + \mu\nu\tau\kappa,
\end{equation}
where  $\kappa = (2\mu-1)/(1-2\nu)$.
The presented theoretical value is therefore calculated for
the above-mentioned transmittances of the beam splitter used.
In order to determine the success probability of the scheme
we measured the coincidence rate of the setup set to perform
MPCC and also the calibration coincidence rate (all the
filters and beam splitter were removed). The ratio of these
two rates determines the success probability calibrated for
``technological losses'' (inherent losses due to
back reflection or systematic error of all the
components)~\cite{Lemr2011}. For more details see
Ref.~\cite{supplement}.

\noindent\textit{Conclusions. } Our implementation presents
a concept of a multifunctional cloner optimized for
quantum communication purposes with respect to {\it a priori}
information about transmitted states and communication
channels. We have experimentally verified the versatile
nature of the proposed cloner. It performs at about 97.5\% of
the theoretical limit for all three regimes tested (UC, PCC,
and MPCC). Thus, in contrast to previous implementations, it
can be used in attacks against a variety of quantum
cryptographic protocols at once~\cite{Gyongyosi}. Some of its
capabilities cannot be provided by previous cloners,
especially for communication through the Pauli damping
channels~\cite{supplement}. Potential applications of our
approach can also include practical quantum networks based on
state-dependent photonic multipliers or amplifiers. We
therefore conclude that this device can be an efficient tool
for a large set of quantum communication and quantum
engineering applications requiring cloning.

\noindent\textit{Acknowledgments.} 
K.L., A.\v{C}., and J.S. gratefully acknowledge the support of the Operational 
Program Research and Development for Innovations of the 
European Regional Development Fund (Project No. 
CZ.1.05/2.1.00/03.0058) and the Operational Program Education
for Competitiveness of the European Social Fund (Project No.
CZ.1.07/2.3.00/20.0017) of the Ministry of Education, Youth 
and Sports of the Czech Republic, from the Institute of Physics
of the Czech Academy of Sciences (Grant no. AVOZ10100522)
and from Palack\'y University (internal Grant No. PrF-2011-009).
A.M. and K.B. acknowledge support from the Polish Ministry of Science
and Higher Education under Grants No. 2619/B/H03/2010/38
and No. 3271/B/H03/2011/40. Support from Grant No.
CZ.1.07/2.3.00/20.0058 is also acknowledged.


\clearpage
\begin{center}
{\Large Supplementary material:}
\end{center}
\section{Optimality proof of the cloners}

\subsection{Average fidelity of cloning a set of qubits}

While considering optimal symmetric $1\to2$ cloning the
optimized figure of merit is the average single-copy
fidelity. We express the average fidelity  as $F={\rm
Tr}\left( R\chi\right)$, where the operator $\chi$ is
isomorphic to a completely-positive trace-preserving
map~\cite{Jamiolkowski72} (CPTP) which performs the cloning
operation and $R=\langle\rho^{\rm T}\otimes(\rho\otimes
1\allowbreak+1\otimes\rho)\rangle/2$, where
$\rho=|\psi\rangle\langle\psi|$ is the density matrix of a
qubit from the cloned set. The matrix $R$ is given explicitly
for any phase-covariant cloner as:
\begin{eqnarray}\label{eq:R}
R=\frac{1}{8}{\left(
\begin{array}{cccccccc}
8c_{2}^{4} & 0 & 0 & 0 & 0 & s_{1}^{2} & s_{1}^{2} & 0 \\
0 & 4c_{2}^{2} & 0 & 0 & 0 & 0 & 0 & s_{1}^{2} \\
0 & 0 & 4c_{2}^{2} & 0 & 0 & 0 & 0 & s_{1}^{2} \\
0 & 0 & 0 & 2s_{1}^{2} & 0 & 0 & 0 & 0 \\
0 & 0 & 0 & 0 & 2s_{1}^{2} & 0 & 0 & 0 \\
s_{1}^{2} & 0 & 0 & 0 & 0 & 4s_{2}^{2} & 0 & 0 \\
s_{1}^{2} & 0 & 0 & 0 & 0 & 0 & 4s_{2}^{2} & 0 \\
0 & s_{1}^{2} & s_{1}^{2} & 0 & 0 & 0 & 0 & 8s_{2}^{4}
\end{array}
\right) },
 \label{N06}
\end{eqnarray}
where $s_i^j=\langle\sin^j(\theta/i)\rangle$ and
$c_i^j=\langle\cos^j(\theta/i)\rangle$ for $i, j=1,2,3...$, and the
angle bracket stands for averaging over the input qubit
distribution $g(\theta,\phi)$. We find the optimal cloning
map $\chi$ by maxi\-mising the functional $F$.

\subsection{The necessary conditions for optimality of the MPCC}

First, let us note that the optimal cloning map $\chi$ must
satisfy the following symmetry conditions imposed by the
symmetry of a mirror phase-covariant set of qubits, i.e.,
$\forall_{\phi\in[0,2\pi]} [R_z(\phi)^*\otimes
R_z(\phi)^{\otimes2},\chi]=0$ and
$[\sigma_x^{\otimes3},\chi]=0$. Second, we assume symmetric
cloning, and therefore we require that both clones have the
same fidelity. Thus, we demand $[\openone_{\rm
in}\otimes{}{\rm SWAP},\chi]=0$. Moreover, we can show that
elements of $\chi$ must be real since maximized fidelity
depends linearly only on the real part of the off-diagonal
elements. We must also remember that $\chi$ must preserve
trace, i.e., ${\rm Tr}_{\rm out}\left[\chi \right]=\openone$.
All the above conditions imply the following form of the map
$\chi$ being a mixture of two CPTP maps:
\begin{eqnarray}\nonumber
\chi &=&(1-p) \left( \begin{array}{cccccccc}
\Lambda^2 & 0 & 0 & 0 & 0& \frac{\Lambda\bar{\Lambda}}{\sqrt{2}}& \frac{\Lambda\bar{\Lambda}}{\sqrt{2}} & 0 \\
0 & \frac{\bar{\Lambda}^2}{2} & \frac{\bar{\Lambda}^2}{2} & 0 & 0 & 0 & 0 &
\frac{\Lambda\bar{\Lambda}}{\sqrt{2}}\\
0 & \frac{\bar{\Lambda}^2}{2} & \frac{\bar{\Lambda}^2}{2} & 0 & 0 & 0 & 0 &
\frac{\Lambda\bar{\Lambda}}{\sqrt{2}}\\
0 & 0 & 0 & 0 & 0 & 0 & 0 & 0\\
0 & 0 & 0 & 0 & 0 & 0 & 0 & 0 \\
\frac{\Lambda\bar{\Lambda}}{\sqrt{2}} & 0 & 0 & 0 & 0 & \frac{\Lambda^2}{2} &
\frac{\Lambda^2}{2} & 0
\\
\frac{\Lambda\bar{\Lambda}}{\sqrt{2}} & 0 & 0 & 0 & 0 & \frac{\Lambda^2}{2} &
\frac{\Lambda^2}{2}& 0
\\
0 & \frac{\Lambda\bar{\Lambda}}{\sqrt{2}}  & \frac{\Lambda\bar{\Lambda}}{\sqrt{2}}  & 0 & 0 & 0 & 0 & \Lambda^2
\end{array} \right)\\&+&p(|011\rangle\langle 011|+|100\rangle\langle 100|).
\end{eqnarray}
where $\bar{\Lambda}^2+\Lambda^2=1$ and $1 \geq \Lambda,~p
\geq 0$ are free parameters. Since we can write the cloning
fidelity as $F = (1-p)F_\Lambda + ps_{1}^{2}/2$ it is
apparent that in order to achieve maximal fidelity assuming
$F > 1/2$ we must set $p=0$. Now, we can derive the
expression for $\Lambda$ demanding that ${\rm d}F/{\rm
d}\Lambda = 0$. As one of the solutions we obtain
\begin{eqnarray}
\Lambda(c^2_1)&=&\sqrt{\frac{1}{2}+\frac{c_1^2}{2\sqrt{P}}},
 \label{N15}
\end{eqnarray}
where $P=2-4c_1^2+3c_1^4$. However, at this point we cannot
conclude that it is optimal.

\subsection{The sufficient conditions for optimality of the MPCC}

As noted by Audenaert and De Moor~\cite{Audenaert02} the
problem of designing an optimal cloning machine can be solved
by means of semidefinite programming. Moreover, it was noted
that as long as $\chi$ is a CPTP map in a convex set, there
are only global extrema. It can be shown that  $\chi$
maximizes fidelity $F={\rm Tr}[R\chi]$ if the following
conditions are satisfied:
\begin{eqnarray}\label{eq:chi1}
(A-R)\chi = 0,\\
\label{eq:chi2}
A - R \geq 0,
\label{lambdaoptimality}
\end{eqnarray}
where $A=\lambda \otimes \openone \geq 0$ is  a positive
semidefinite matrix of Lagrange multipliers ensuring that
$\chi$ is CPTP  map, i.e., ${\rm Tr}_{\rm
out}(\chi)=\openone_{\rm in}$. The operator  $\lambda={\rm
Tr}_{\rm out}(R\chi)$ is derived by demanding that the
variance of fidelity $F$ over $\chi$ should be equal to zero.
If the condition (\ref{eq:chi1}) is satisfied, then for any
CPTP map $\chi$ we obtain $\mathrm {Tr}[(A -R)\chi] \geq 0$.
It also follows from the trace preservation condition that
${\rm Tr}(A\chi)={\rm Tr}\lambda$. Hence, the fidelity $F$ is
bounded by the trace preservation condition and $F \leq {\rm
Tr}[\lambda]$. If inequality is saturated by $\chi$, then
$\chi$ represents the optimal cloning transformation.

For MPCC we have
\begin{equation} \label{N20}
\lambda=\frac{1}{4}[(1+c_1^2)\Lambda^2 +\bar{\Lambda}^2+\sqrt{2}(1-c_1^2)\Lambda\bar{\Lambda}]\,\openone_{\rm in}.
\end{equation}
Henceforth, it can be easily shown that ${\rm Tr}\lambda-F=0$.
The eigenvalues of operator $\Delta=A - R$ can be expressed in terms of $R$ matrix elements in the following way:
\begin{eqnarray}
\nonumber
\delta_1&=&\frac{1}{2}\left(F-\frac{1}{2}\right),\\\nonumber
\delta_2&=&\frac{1}{2}\left(F-\frac{1-c_1^2}{2}\right),\\
\delta_{3,4}&=&\frac{1}{2}\left(F-R_{1,1}-R_{2,2}\pm\bar
R\right),\label{N22a}
\end{eqnarray}
where $\bar R^2=(R_{1,1}-R_{2,2})^2+8R_{1,6}^2$. All the eigenvalues are
double degenerated. Moreover, we have
\begin{equation}\label{N23}
F=R_{1,1}+R_{2,2}+\bar R.
\end{equation}
Thus, $\delta_3=F-(2+c_1^2)/4$ and $\delta_4=0$. Since
$F>3/4$, $\forall_i \delta_i\ge0$, we conclude that $\Delta$
is a positive semidefinite matrix. Thus, we have shown that
the conditions~(\ref{eq:chi1}) and~(\ref{eq:chi2}) are
satisfied, which completes the proof.

\section{Compensating for imperfect transmittances}
In our case the equation relating beam-splitter
transmittances ($\mu + \nu = 1$) does not hold and we have
$\mu + \nu \neq 1$. Hence additional filtering operations are
required in order to maintain the maximum achievable fidelity
of the setup. This additional filtering manifests itself in
two ways. First, one needs to unbalance the ancilla-dependent
filtering performed by filters F in both BDAs. We require
$\tau_1 = \tau$ and $\tau_2 = \omega\tau$ for the BDA1 and
BDA2, respectively, where
\begin{equation}
\omega=\frac{\tau_2}{\tau_1}={\frac{\mu\nu}{(1-\mu)(1-\nu)}}.
\label{eq:omega}
\end{equation}
Note that $\omega=1$ in the ideal case for $\mu+\nu=1$ and
$\omega = 0.726$ for the applied PDBS. Second, the
realization of the MPCC with the PDBS where $\mu+\nu\neq1$
requires applying an additional unconditional filtering. This
filtering is polarization-dependent and is performed
regardless of the state of the ancillary photon. The
polarization dependent transmittances $\tau_H$ and $\tau_V$
for the $H$ and $V$-polarized photons, respectively, need to
satisfy the following relation:
\begin{equation}
\kappa = \frac{\tau_V}{\tau_H} = {\frac{2\mu-1}{1-2\nu}},
\label{eq:kappa}
\end{equation}
where $\kappa$ ($\kappa=1$ for $\mu+\nu=1$) is a constant
value fixed by the parameters of the PDBS (in our case
$\kappa = 0.838$), and both $\tau_H$ and $\tau_V$ should have
the largest possible values in order to maximize the
efficiency of the setup. Therefore, we apply an additional
unconditional filtering only for the $V$-polarized photons
since the optimal transmittances are $\tau_V = \kappa$ and
$\tau_H = 1$. Please note that for our PDBS $\kappa<1$ and in
the opposite case the best choice of the parameters is
$\tau_V = 1$ and $\tau_H = 1/\kappa$. Moreover, if there are
any other systematic uniform polarization-dependent losses
$\tau_H^\prime$ and $\tau_V^\prime$ we can compensate for
them by setting $\kappa = \tau_H^\prime/\tau_V^\prime\times
\left({2\mu-1)/(1-2\nu}\right)$.

To summarize the above-mentioned corrections, the overall
filtering operations in the first mode are described by
\begin{eqnarray}
\tau_{1,H} = \tau^{\delta_{V,s}} \mbox{ and }\tau_{1,V} = \kappa\tau^{\delta_{H,s}},
\label{eq:filter1}
\end{eqnarray}
and in the second mode by
\begin{eqnarray}
\tau_{2,H} = (\omega\tau)^{\delta_{V,s}} \mbox{ and } \tau_{2,V} = \kappa(\omega\tau)^{\delta_{H,s}},
\label{eq:filter2}
\end{eqnarray}
where $\delta_{V,s}$ ($\delta_{H,s}$) is Kronecker's delta
and is equal to 1 iff the polarization $s$ of the ancillary
photon is $V$ ($H$).

To implement the required filtering, additional
polarization-independent filters FA1 and FA2 are placed at
the output modes. These two filters together with the filters
in both BDAs are sufficient to perform filtering operation
described by Eqs.~(\ref{eq:filter1}) and~(\ref{eq:filter2}).

The usage of additional filtering saves the maximum
achievable fidelity at the expense of lowering the success
probability of the scheme. Using  PDBA transmittances and the
parameter $\Lambda$ of the cloned state one can express the
expected success probability of the scheme in the form of
\begin{equation}
P_{\rm th} =  (1-2\mu)^2/2 + \mu\nu\tau\kappa,
\label{eq:psucc}
\end{equation}
where  $\kappa = (2\mu-1)/(1-2\nu)$. \\

\section{Implementing the generalized PCC and axisymmetric cloning}

By using the same setup we implemented the generalized PCC (see Tab.~\ref{tab:pcc})
which is a special case of the axisymmetric cloner described in Ref.~\cite{Karol10}.
In order to perform arbitrary axisymmetric cloning we set parameters of filters according to
the following relations:
\begin{equation}
\begin{cases}
\frac{\tau_{1,H}}{\tau_{1,V}}=\left(\frac{\cos\alpha_{+}}{\sin\alpha_{-}}\right)^{2}\frac{2\left(1-\mu\right)\left(1-\nu\right)}{\left(1-2\mu\right)^{2}}\\
\frac{\tau_{2,H}}{\tau_{2,V}}=\left(\frac{\cos\alpha_{+}}{\sin\alpha_{-}}\right)^{2}\frac{2\mu\nu}{\left(1-2\mu\right)^{2}}
\end{cases}\mathrm{for}\: s=H,
\end{equation}
and
\begin{equation}
\begin{cases}
\frac{\tau_{1,H}}{\tau_{1,V}}=\left(\frac{\sin\alpha_{+}}{\cos\alpha_{-}}\right)^{2}\frac{\left(2\nu-1\right)^{2}}{2\left(1-\mu\right)\left(1-\nu\right)}\,\\
\frac{\tau_{2,H}}{\tau_{2,V}}=\left(\frac{\sin\alpha_{+}}{\cos\alpha_{-}}\right)^{2}\frac{\left(2\nu-1\right)^{2}}{2\mu\nu}
\end{cases}\mathrm{for}\: s=V,
\end{equation}
where $\alpha_{\pm}$ is given by Eq.~(14) from \cite{Karol10} and
$s=H,\, V$ stands for polarization of the ancillary state. For the
generalized PCC we picked  $s$ deterministically. We set 
$\alpha_{+}=\pi/2$ and $\alpha_{-}=0$ for $s=H$ when
we cloned a qubit from the northern hemisphere, alternatively we set
$\alpha_{+}=0$ and $\alpha_{-}=\pi/2$ for $s=V$ each time when
the cloned qubit was from the southern hemisphere. Please note that
we could have also picked ancillary state at random (as for the MPCC),
but then we would have had to block all the output modes for half
of the cases.

\section{Measuring the success probability}

In order to measure success probability we need to estimate
the inherent technological losses of the scheme and the
initial photonic rate. The technological losses occur as a
result of detector efficiencies, fiber coupling losses or
back reflections. The coincidence rate $C_{\rm clon}$
measured at the end of the working cloner can be expressed as

\begin{equation}
C_{\rm clon} = P_{\rm ex}  \tau_{\rm tech}  C_{\rm init},
\label{eq:cc_rate}
\end{equation}
where $P_{\rm ex}$ denotes the success probability of the
cloning scheme, $\tau_{\rm tech}$ denotes the transmittance
of the setup due to  technological losses and $C_{\rm init}$
is the initial rate of photon pairs from the source. To
compensate for the technological losses and the initial
photon rate we use the following calibration procedure:  PDBS
is placed on a translation state allowing us to shift it
slightly so that the reflected beam is no longer coupled. We
use $|H_1H_2\rangle$ for the input state knowing that the
beam splitter would decrease the coincidence rate by the
factor of $1/\mu^2$. In this configuration we remove all the
neutral density filters and measure the calibration
coincidence rate $C_{\rm calib}$ at the end of the scheme.
One can clearly see that
\begin{equation}
C_{\rm calib} = \mu^2  \tau_{\rm tech}  C_{\rm init}
\label{eq:calib}
\end{equation}
so the success probability of the cloning operation can be
expressed by combining Eqs.~(\ref{eq:cc_rate})
and~(\ref{eq:calib}):
\begin{equation}
P_{\rm ex} = \mu^2 \frac{C_{\rm clon}}{C_{\rm calib}}.
\label{eq:pex}
\end{equation}
This equation allows us to obtain the success probability of
the cloning operation from the measurement of two coincidence
rates: the first is the coincidence rate of the working
cloner and the second is the calibration coincidence rate.
Note that Eqs.~(\ref{eq:psucc}) and~(\ref{eq:pex}) describe
the same quantity.

\section{Measured values}
\begin{figure}
\includegraphics[scale=0.9]{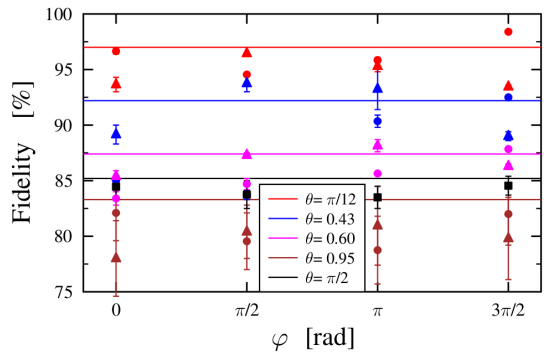}
\caption{\label{fig:phase} Phase (angle
$\varphi$) dependence of fidelity of the MPCC for the selected values
(see Table~\ref{tab:mpcc}) of angle $\theta$.}
\end{figure}

Our detailed summary of measured and predicted results is
presented in Tables~\ref{tab:pcc},~\ref{tab:mpcc},
and~\ref{tab:uc}. In Fig.~\ref{fig:phase} we show how 
the cloning fidelity of the MPCC varies with phase $\varphi$.
\begin{table}
\begin{ruledtabular}
\begin{tabular}{lll}
Angle $\theta$  &   $F_{\rm ex}$ [\%]    & $F_{\rm th}$ [\%]  \\\hline
$0$             &   99.8 $\pm$ 0.4          & 100.0                 \\\hline
$\pi/12$        &   99.3 $\pm$ 0.4          & 99.8                  \\\hline
$\pi/5$         &   98.0 $\pm$ 0.8          & 98.8                  \\\hline
$\pi/3$         &   95.7 $\pm$ 0.8          & 95.3                  \\\hline
$3\pi/8$        &   92.4 $\pm$ 1.5          & 93.4                  \\\hline
$\pi/2.25$      &   88.7 $\pm$ 1.1          & 89.4                  \\\hline
$\pi/2$         &   84.1 $\pm$ 0.5          & 85.4                  \\\hline
$\pi/1.8$       &   87.9 $\pm$ 0.7          & 89.4                  \\\hline
$5\pi/8$        &   91.3 $\pm$ 1.0          & 93.4                  \\\hline
$2\pi/3$        &   95.0 $\pm$ 0.8          & 95.3                  \\\hline
$4\pi/5$        &   97.9 $\pm$ 0.7          & 98.8                  \\\hline
$11\pi/12$      &   98.4 $\pm$ 1.0          & 99.8                  \\\hline
$\pi$           &   99.8 $\pm$ 0.4          & 100.0                 \\
\end{tabular}
\end{ruledtabular}
\caption{\label{tab:pcc}Summarized data for the PCC regime.
$F_{\rm ex}$ denotes experimentally estimated average
fidelity for a  given polar angle $\theta$ on the Bloch
sphere and $F_{\rm th}$ is the theoretical prediction. Note
that the error estimated as RMS is just indicative of the
actual error, because it does not take into account the
physical properties of fidelity.}
\end{table}

\begin{table}
\begin{ruledtabular}
\begin{tabular}{lllll}
Angle $\theta$  &   $F_{\rm ex}$ [\%]    & $F_{\rm th}$ [\%]  & $P_{\rm ex}$ [\%]  & $P_{\rm th}$ [\%]  \\\hline
$0$ & 99.6 $\pm$ 0.4 & 100.0 & 10.5 $\pm$ 2.8 & 13.3\\\hline
$\pi/12$ & 95.6 $\pm$ 1.7 & 97.0 & 10.6 $\pm$ 1.9 & 13.3\\\hline
0.43 & 89.6 $\pm$ 0.4 & 90.4 & 10.4 $\pm$ 1.5 & 13.5\\\hline
$\pi/5$ & 86.1 $\pm$ 1.6 & 87.4 & 9.6 $\pm$ 0.9 & 14.3\\\hline
$\pi/4$ & 81.9 $\pm$ 2.0 & 84.1 & 14.0 $\pm$ 2.9 & 16.2\\\hline
0.95 & 80.2 $\pm$ 1.5 & 83.3 & 19.5 $\pm$ 3.5 & 18.6\\\hline
$3\pi/8$ & 82.3 $\pm$ 1.3 & 84.0 & 23.7 $\pm$ 1.5 & 21.7\\\hline
$\pi/2$ & 84.1 $\pm$ 0.5 & 85.4  & 24.8 $\pm$ 0.1 & 24.0\\\hline
$5\pi/8$ & 82.3 $\pm$ 1.3 & 84.0 & 23.7 $\pm$ 1.5 & 21.7\\\hline
$2\pi/3$ & 80.2 $\pm$ 1.5 & 83.3  & 19.5 $\pm$ 3.5 & 18.6\\\hline
2.19 & 81.9 $\pm$ 2.0  & 84.1 & 14.0 $\pm$ 2.9  & 16.2\\\hline
$4\pi/5$ & 86.1 $\pm$ 1.6  & 87.4 & 9.6 $\pm$ 0.9 & 14.3\\\hline
2.71 & 89.6 $\pm$ 0.4 & 90.4 & 10.4 $\pm$ 1.5 & 13.5\\\hline
$11\pi/12$ & 95.6 $\pm$ 1.7 & 97.0 & 10.6 $\pm$ 1.9  & 13.3\\\hline
$\pi$ & 99.6 $\pm$ 0.4 & 100.0 & 10.5 $\pm$ 2.8 & 13.3\\
\end{tabular}
\end{ruledtabular}
\caption{\label{tab:mpcc}Same as in Table~\ref{tab:pcc} but
for the MPCC regime. Moreover $P_{\rm ex}$ and $P_{\rm th}$
denote experimental and theoretical success probabilities.}
\end{table}

\begin{table}
\begin{ruledtabular}
\begin{tabular}{lll}
Polarization state  &   $F_{\rm ex}$ [\%]    & $F_{\rm th}$ [\%]  \\\hline
horizontal              &   80.2 $\pm$ 3.1          & 83.3                  \\\hline
diagonal        &   81.5 $\pm$ 1.5          & 83.3                  \\\hline
anti-diagonal           &   81.3 $\pm$ 0.2          & 83.3                  \\\hline
right-circular          &   82.5 $\pm$ 1.4          & 83.3                  \\\hline
left-circular       &   80.1 $\pm$ 0.9          & 83.3                  \\\hline
vertical        &   83.2 $\pm$ 0.3          & 83.3                  \\
\end{tabular}
\end{ruledtabular}
\caption{\label{tab:uc}Same as in Tables~\ref{tab:pcc} and~\ref{tab:mpcc}
but for the UC regime and various polarization states.}
\end{table}

\section{The MPCC and Pauli damping channel}

Here, we show that our implementation of the MPCC can be
interpreted as a quantum simulation of the Pauli dampening
channel, where an error (bit-flip error, phase-flip error or
both) occurs with some probability. This correspondence can
lead to immediate applications of the proposed device for
quantum eavesdropping. The density matrices of both clones
are the same as the density matrix of the copied state
transmitted via the noisy channel,
$$
\hat{\rho}_{\rm out} = \alpha_+\hat{\rho}_{\rm in} +\frac{
\bar{\Lambda}^2}{4}(\hat{\sigma}_x\hat{\rho}_{\rm
in}\hat{\sigma}_x +\hat{\sigma}_y\hat{\rho}_{\rm
in}\hat{\sigma}_y)+\alpha_-\hat{\sigma}_z\hat{\rho}_{\rm
in}\hat{\sigma}_z,
$$
where
$\alpha_{\pm}=\left({1+\Lambda^2}\pm2{\sqrt{2}}\Lambda\bar{\Lambda}\right)/4$
and $\Lambda^2 + \bar{\Lambda}^2 = 1$. The parameter
$\Lambda$ depends on the distribution $g$ of the cloned
qubits and $\hat{\rho}_{\rm in} = \ket{\psi}\bra{\psi}$. In
the special case for $\Lambda^2 = 2/3$, the channel becomes
so-called depolarizing channel, where the probability of all
errors is the same and equal to 1/12. In such case the
corresponding cloning machine is the UC. Moreover, for
$\Lambda^2 = 1$ the channel becomes a dephasing channel (only
the phase-flip error can occur) and the corresponding cloning
is optimized for covariant cloning of the eigenstates of the
phase-flip operator $\ket{0}$ and $\ket{1}$ and, thus, those
two states can be perfectly copied or transmitted through the
lossy channel.


\begin{thebibliography}{99}
\bibitem{cryptography1}

G.~Van~Assche, {\it Quantum Cryptography and Secret-Key
Distillation} (Cambridge University Press, Cambridge, 2006).

\bibitem{Bruss11}
{\it Lectures on Quantum Information}, edited by
 D.~Bru\ss{} and G.~Leuchs  (Wiley-VCH, Berlin,
2011).

\bibitem{Buzek96}
V.~Bu\v{z}ek and M.~Hillery, \extra{Quantum copying:
Beyond the no-cloning theorem,} \pra~\textbf{54}, 1844
(1996).

\bibitem{cloning1}
V.~Scarani, S.~Iblisdir, N.~Gisin, and A.~Acin,
\extra{Quantum cloning,}  {Rev. Mod. Phys.}~{\bf 77},
1225 (2005).

\bibitem{cloning2}
N.~J. Cerf and J.  Fiur\'a\v{s}ek, \extra{Optical quantum
cloning,} {in \it Progress in Optics}, edited by E. Wolf (Elsevier,
Amsterdam, 2006), Vol. 49, p. 455.

\bibitem{Ricci04}
M. Ricci, F.~Sciarrino, C.~Sias, and F.~De~Martini,
\extra{Teleportation Scheme Implementing the Universal
Optimal Quantum Cloning Machine and the Universal NOT Gate,}
\prl~\textbf{92}, 047901 (2004).

\bibitem{Irvine04}
W.~T.~M. Irvine, A.~L.~Linares, M.~J.~A.~de~Dood, and 
D.~Bouwmeester, \prl~\textbf{92}, 047902 (2004).

\bibitem{Khan04}
I.~Ali~Khan and J.C.~Howell,
\extra{Hong-Ou-Mandel cloning:
Quantum copying without an ancilla,} \pra~\textbf{70},
010303(R) (2004).

\bibitem{cloning3}
A. \v{C}ernoch, L.~Bart{\accent23 u}\v{s}kov\'{a}, J.~Soubusta, M.~Je\v{z}ek,
J.~Fiur\'{a}\v{s}ek, M.~\& Du\v{s}ek,
\extra{Experimental phase-covariant cloning of
polarization states of single photons,}  \pra~ \textbf{74},
042327 (2006).

\bibitem{cloning4}
H.~Chen, X.~Zhou, D.~Suter, and J.~Du,
\extra{Experimental
realization of $\mathbf{1}\mathbf{\rightarrow}\mathbf{2}$
asymmetric phase-covariant quantum cloning,}  \pra {\bf 75},
012317 (2007).

\bibitem{cloning5}
L.~Bart{\accent23 u}\v{s}kov\'{a}, M.~Du\v{s}ek, A.~\v{C}ernoch, J.~Soubusta, and
J.~Fiur\'a\v{s}ek,
\extra{Fiber-optics implementation of an asymmetric
phase-covariant quantum cloner,} \prl~\textbf{99}, 120505
(2007).

\bibitem{cloning6}
J.~Soubusta, L.~Bart\ifmmode \mathring{u}\else \r{u}\fi{}\ifmmode \check{s}\else \v{s}\fi{}kov\'a, A.~\ifmmode \check{C}\else \v{C}\fi{}ernoch,  M.~Du\ifmmode \check{s}\else  \v{s}\fi{}ek, and J.~Fiur\'a\ifmmode \check{s}\else \v{s}\fi{}ek,
\extra{Experimental
asymmetric phase-covariant quantum cloning of polarization
qubits,} \pra~{\bf 78}, 052323 (2008).

\bibitem{Bruss00}
D.~Bru\ss{}, M.~Cinchetti, G.~M.~ D'Ariano, and C.~Macchiavello,
\extra{Phase-covariant quantum cloning,}
\pra~\textbf{62}, 012302 (2000).

\bibitem{DAriano03}
G.~M.~D'Ariano and C.~Macchiavello,
\extra{Optimal phase-covariant cloning for qubits and qutrits,}
\pra~\textbf{67}, 042306 (2003).

\bibitem{F2}
J.~Fiur\'a\v{s}ek, \extra{Optical implementations of the
optimal phase-covariant quantum cloning machine,}
\pra~\textbf{67}, 052314 (2003).

\bibitem{Bartkiewicz2}
K.~Bartkiewicz and A.~Miranowicz,
 \extra{Optimal cloning
of qubits given by an arbitrary axisymmetric distribution on
the Bloch sphere,} \pra~\textbf{82}, 042330 (2010).

\bibitem{Bartkiewicz}
K.~Bartkiewicz, A.~Miranowicz, and \c{S}.~K.~\"Ozdemir,
\extra{Optimal mirror phase-covariant cloning,}
\pra~\textbf{80}, 032306 (2009).

\bibitem{Lemr2011}
K.~Lemr, A.~\ifmmode \check{C}\else \v{C}\fi{}ernoch, J.~Soubusta, K.~Kieling,
Eisert,~J., and M.~\& Du\ifmmode \check{s}\else \v{s}\fi{}ek,
\extra{Experimental
Implementation of the Optimal Linear-Optical Controlled Phase
Gate,} \prl~\bf 106\normalfont , 013602 (2011).

\bibitem{Halenkova12}
E.~Halenkov\'{a}, A.~\v{C}ernoch, K.~Lemr, J.~Soubusta, S.~Drusov\'{a},
App. Optics, {\bf 51}(4), 474 (2012).

\bibitem{Jezek03}
M.~Je\ifmmode \check{z}\else \v{z}\fi{}ek, J.~Fiur\'a\ifmmode \check{s}\else \v{s}\fi{}ek, and Z.~Hradil,
\extra{Quantum inference of states and processes,} \pra~{\bf
68}, 012305 (2003).

\bibitem{Gyongyosi}
L.Gyongyosi and S. Imre,
\extra{Efficient Computational
Information Geometric Analysis of Physically Allowed Quantum
Cloning Attacks for Quantum Key Distribution Protocols,}
{WSEAS Trans. Commun.} {\bf 9}, 165 (2010).

\bibitem{supplement}
See supplementary material below
for more detailed
information on technical aspects of the experimental setup,
optimality proof of the cloners and Pauli damping channels.

\end{thebibliography}

\begin{thebibliography}{99}

\bibitem{Jamiolkowski72}
A. Jamio\l{}kowski, \extra{Linear transformations which
preserve trace and positive semidefiniteness of operators,}
Rep. Math. Phys. \textbf{3}, 275 (1972).

\bibitem{Audenaert02}
K. Audenaert and B. De Moor, \extra{Optimizing completely
positive maps using semidefinite programming,} Phys. Rev. A
\textbf{65}, 030302 (2002).

\bibitem{Karol10}
K. Bartkiewicz and A. Miranowicz,
 \extra{Optimal cloning
of qubits given by an arbitrary axisymmetric distribution on
the Bloch sphere,} \pra~\textbf{82}, 042330 (2010).

\end{thebibliography}
\end{document}